\documentclass[preprint,aps,prd,showpacs,nofootinbib]{revtex4}
\usepackage{graphicx}
\usepackage{amsmath}
\usepackage{color}
\usepackage{bm}

\newcommand{\be}{\begin{equation}}
\newcommand{\ee}{\end{equation}}
\newcommand{\bea}{\begin{eqnarray}}
\newcommand{\eea}{\end{eqnarray}}

\newcommand{\mt}[1]{\textrm{\tiny #1}}

\newcommand{\rh}{r_\mt{H}}

\begin{document}

\begin{center}
{\bf Sound modes and stability of momentum dissipated black branes in holography}\\

\vspace{1.6cm}

Wenhe Cai $^{~1}$, Xian-Hui Ge $^{~1,2,*}$\let\thefootnote\relax\footnotetext{* Corresponding author. gexh@shu.edu.cn},~~~  QingBing Wang  $^{~1}$ \\
\vspace{0.8cm}

$^1${\it Department of Physics, Shanghai University, Shanghai 200444,  China} \\
$^2${\it Center for Gravitation and Cosmology, College of Physical Science and Technology,
Yangzhou University, Yangzhou 225009, China}
\vspace{1.6cm}

\begin{abstract}
  We systematically investigate the sound modes of momentum dissipated holographic systems. In particular, we focus on the Einstein-linear axions and the Einstein-Maxwell-dilaton-axion theories in four-dimensional bulk spacetime dimensions. The sound velocities of the two theories are computed respectively and the sound attenuation of the Einstein-Maxwell-axion theory is also calculated analytically. We also obtain numeral dispersion relations in the two theories which match with our analytical results. Our results show that the sound velocity of the Einstein-Maxwell-dilaton theory with additional linear axion fields is equivalent to that of 2 + 1 - dimensional Banados-Teitelboim-Zanelli black holes. It allowed us to compare our solution of the Einstein-linear axions theory with that of systems without translational invariance from another method. After the computation on the sound velocity, we calculate the quasinormal modes of scalar-type fluctuations in the Einstein-Maxwell-dilaton-axion theory. The results show that a dynamical instability is observed under the condition that the null energy condition is violated.
\end{abstract}
\end{center}
\maketitle
\section{Introduction}
The AdS/CFT correspondence provides a useful tool for the computation of the transport coefficient of strongly coupled systems by analyzing small perturbations of the black branes near their equilibrium states \cite{maldacena,gubser,witten}.
For realistic materials without momentum conservation because of impurities and lattices, the translational symmetry breaking black brane solution must be employed \cite{withers,massive}.
The thermoelectric conductivities and magnetotransport coefficients, which are mainly related to the vector-type perturbation of the metric, are widely investigated in those models. For tensor type perturbation, the shear viscosity is relevant (see \cite{poov,jah,hartnoll,wang,ling} for related references). Any finite temperature medium should propagate sound, the computation of sound velocity, the quasinormal modes of the scalar-type metric perturbation, and its dispersion relation in momentum dissipated systems are important for a complete understanding of holography.

In this paper, we will study the sound mode and the related dynamical stability of those momentum dissipated black holes. In addition to the computation of sound velocity, we will also study quasinormal modes, which corresponds to the scalar type perturbation of such black holes. The main content of our paper consists of two parts. In the first part, we focus on Einstein-linear axion model. Actually, sound modes in this model have been investigated in the previous work \cite{Davison:2014lua}. Two sound-like poles collide and produce two purely imaginary poles when the decay rate $\Gamma$ is approximately equal to the momentum $k$ (i. e. $\Gamma\sim k$), through the analysis in \cite{Davison:2014lua}. However, we provide another analytical method in this paper. In the second part, we focus on Einstein-Maxwell-dilaton-axion model. In this model, the sound mode has not been studied yet. In \cite{Ge:2016sel}, a black hole solution in the asymptotic Lifshitz spacetime with a hyperscaling violating factor was utilized to compute the DC thermoelectric conductivities analytically. Both the linear-T and the quadratic-T contributions to the resistivity were realized in that model \cite{Ge:2016sel}. But the classical null energy condition is violated.
 Another motivation of this paper comes from the fact that the energy driven by the accelerated expansion of the Universe is a kind of energy with large negative pressure, termed dark energy. The equation-of-state parameter for dark energy $w$
 can be smaller than $-1$. This seems not only admissible but even preferable for describing an increasing acceleration as from the most recent estimates $w=-1.03\pm 0.03$, combining with Type Ia supernova and baryon acoustic oscillations data \cite{Aghanim:2018eyx}. Since such fields may be important in our Universe, it is interesting to check their behavior in local phenomena, for example, in holographic black holes.
 Here, we will study stability of the black branes where the null energy condition is violated. Accordingly, we check the imaginary part of the quasinormal modes for the sound channel. Some unstable quasinormal modes correspond to the instability of these momentum dissipated theories.

The paper is organized as follows: In sec. II, as a test example, we consider the Einstein-linear axion model, with the sign of the linear axion terms to be $+1$ and $-1$. The negative sign corresponds to a phantom-like energy. We focus on the black brane system in the presence of the linear axions. We derive the speed of sound and the sound attenuation analytically. Numerical computations of the quasinormal modes of such scalar-type perturbations are then carried out.  In Sec. III, we review the Einstein-Maxwell-dilaton theory with additional axion fields. In Sec. IV, the tensor- type perturbation and the boundary field theory are investigated. In Sec. V, we start to construct the sound channel. We evaluate the speed of sound, the sound-like quasinormal spectrum with and without the phantom-like term. The Sec. VI is the summary and discussion. In the appendix, we collect the master equations.

\section{Sound modes of black brane with linear axions}
In this section, let us focus on an Einstein-axion theory of momentum relaxation only consisting of linear axion fields \cite{withers}. This model plus a Maxwell field is dual to a simple translational symmetry breaking condensed matter system and has been widely studied recently.   The action is given as
\bea
S&=&\int_M \! d^{4}x \!\sqrt{-g} \bigg[\bigg(R-2\Lambda-\frac{\varepsilon}{2}\!\sum^{2}_{i=1}\partial\chi^2_{i}\bigg)\bigg]
-2\int_{\partial M} \! d^{3}x \!\sqrt{-\gamma}K,
\eea
where $\Lambda=-3$ and $\varepsilon=\pm 1$ distinguishes normal, canonical scalar fields ($\varepsilon=+1$) and
phantom fields ($\varepsilon=-1$). The black brane solution takes the form  \cite{withers}
\bea
ds^2&=&-f(r)dt^2+\frac{dr^2}{f(r)}+r^2(dx^2+dy^2)\,,\nonumber\\
f(r)&=&r^2-\varepsilon\frac{\beta}{2}-\frac{m}{r},~~~
  \chi_i= \beta \delta_{ia} x^a,
\eea
where $a$ labels the spatial directions $\{x,y\}$, $i$ labels the axion fields and $\beta$ is a real-valued constant. In the absence of the linear axion fields, the metric reduces to three-dimensional Schwarzschild-anti-de Sitter (AdS) black hole. The Hawking temperature of this black hole is given by $T=\frac{1}{4\pi}(3\rh-\frac{\beta^2}{2\rh})$ and the entropy density $s=4\pi \rh^2 $ where $\rh$ is the horizon radius.

The tensor mode of this black hole and the shear viscosity have been studied in \cite{hartnoll} and the shear viscosity bound was found to be violated strongly. The vector mode and corresponding electrical conductivity are also studied in \cite{withers}. For the sound mode, we are going to choose the following gauge as $h_{\mu r}=0$ \footnote{After careful calculations, we are confirmed that the gauge $h_{\mu r}=0$ is consistent with the whole equations of motion even when the linear axions  break the translational invariance.}. In the absence of the Maxwell field, the only nonzero fluctuations are $h_{tt}$, $h_{xx}$, $h_{yy}$, $h_{tx}$ and $\delta \chi_i$ (see also \cite{Bertoldi:2009yi}).
We focus on a single Fourier component that propagates along the $x$ direction
\bea
h_{tt}=-e^{-i\omega t+i kx}g_{tt}H_{tt}&,& h_{tx}=h_{xt}=e^{-i\omega t+i kx}g_{xx}H_{tx},\\
h_{xx}=e^{-i\omega t+i kx}g_{xx}H_{xx}&,& h_{yy}=e^{-i\omega t+i kx}g_{yy}H_{yy},\\
\delta \chi_i=e^{-i\omega t+i kx}\mathcal{X}_i (r)&.&
\eea
 The linearized equations of motion for the nonzero fluctuations $h_{tt}$, $h_{xx}$, $h_{yy}$, $h_{tx}$ and $\delta \chi_i$ could be obtained as,
\bea
&&H''_{tt}+\ln'\bigg(\frac{g_{tt}g_{xx}}{g^{1/2}_{rr}}\bigg)H'_{tt}-\ln'(g^{1/2}_{tt})H'_{ii}-g_{rr}\bigg(\frac{\omega^2}{g_{tt}}H_{ii}+\frac{k^2}{g_{xx}}H_{tt}+2\frac{\omega k}{g_{tt}}H_{tx}\bigg)=0\,,\label{f1}\\
&&H''_{tx}+\ln'\bigg(\frac{g^2_{xx}}{g^{1/2}_{rr}g^{1/2}_{tt}}\bigg)H'_{tx}+\frac{g_{rr}}{g_{xx}}\omega k H_{aa}+h(\beta)H_{tx}-i\varepsilon\beta\omega \frac{g_{rr}}{g_{xx}}\mathcal{X}_1=0\,,\label{f2}\\
&& H''_{xx}+\ln'\bigg(\frac{g^{1/2}_{tt}g^{3/2}_{xx}}{g^{1/2}_{rr}}\bigg)H'_{xx}+\ln'(g^{1/2}_{xx})\bigg(H'_{yy}-H'_{tt}\bigg)+h(\beta)H_{xx}\nonumber\\
&&+g_{rr}\bigg(-\frac{\omega^2}{g_{tt}}H_{xx}+\frac{k^2}{g_{xx}}(H_{tt}-H_{yy})+2\frac{\omega k}{g_{tt}}H_{tx}\bigg)+2i\varepsilon k\beta\frac{g_{rr}}{g_{xx}}\mathcal{X}_1=0\,,\label{f3}\\
&& H''_{yy}+\ln'\bigg(\frac{g^{1/2}_{tt}g^{3/2}_{xx}}{g^{1/2}_{rr}}\bigg)H'_{yy}+\ln'(g^{1/2}_{xx})\bigg(H'_{xx}-H'_{tt}\bigg)-g_{rr}\bigg(\frac{\omega^2}{g_{tt}}+\frac{k^2}{g_{xx}}\bigg)H_{yy}\nonumber\\
&&+h(\beta)H_{yy}=0\,,\label{f4}\\
&&\mathcal{X}''_1+\bigg(\ln\frac{g^{1/2}_{tt}g_{xx}}{g^{1/2}_{rr}}\bigg)\mathcal{X}'_1+g_{rr}\bigg(\frac{\omega^2}{g_{tt}}-\frac{k^2}{g_{xx}}\bigg)\mathcal{X}_1
\nonumber\\
&&-i\beta g_{rr}\bigg[\frac{k}{2g_{xx}}H_{tt}+\frac{\omega}{g_{tt}}H_{tx}+\frac{k}{2 g_{xx}}(H_{xx}-H_{yy})\bigg]=0\,,\label{f5}
\eea
where $H_{ii}=\{H_{xx}, H_{yy}\}$ and
\be
 h(\beta)=2\Lambda g_{rr}+\frac{g'_{tt}g'_{xx}}{2g_{tt}g_{xx}}-\frac{g'_{rr}g'_{xx}}{2g_{rr}g_{xx}}+\frac{g''_{xx}}{g_{xx}}
\ee
Besides, there are three additional first-order constraint  equations associated with the gauge fixing condition given as,
\bea
&&H'_{ii}+\ln'\bigg(\frac{g^{1/2}_{xx}}{g^{1/2}_{tt}}\bigg)H_{ii}+\frac{k}{\omega}H'_{tx}+2\frac{k}{\omega}\ln'\bigg(\frac{g^{1/2}_{xx}}{g^{1/2}_{tt}}\bigg)H_{tx}=0\,,\nonumber\\
&&H'_{tt}-\ln'\bigg(\frac{g^{1/2}_{xx}}{g^{1/2}_{tt}}\bigg)H_{tt}+\frac{\omega}{k}\frac{g_{xx}}{g_{tt}}H'_{tx}-H'_{yy}+i\varepsilon\frac{\beta}{q}\chi'_1=0\,,\\
&&\ln'\bigg(g^{1/2}_{xx}g^{1/2}_{tt}\bigg)H'_{ii}-\ln'(g_{xx})H'_{tt}+g_{rr}\bigg[-\frac{\omega^2}{g_{tt}}H_{ii}+\frac{k^2}{g_{xx}}(H_{tt}-H_{yy})+2\frac{\omega k}{g_{tt}}H_{tx}\bigg]
\nonumber\\
&&+i\varepsilon\beta k\frac{g_{rr}}{g_{xx}}\mathcal{X}_1+\frac{h(\beta)}{2}H_{ii}=0\,.
\eea
According to the incoming boundary conditions at $u=1$ and the Dirichlet boundary conditions at $u=0$, we solve the following differential equations for $\varepsilon=\pm1$
 \begin{align}
   &\frac{3 f'H'_{tt}}{2 f}-\left(\frac{2}{r}+\frac{f'}{2 f}\right) \left(H'_{xx}+H'_{yy}\right)+H''_{tt}-\left(H''_{xx}+H''_{yy}\right)=0\,,\\
   &k\left[H_{tt} \left(\frac{1}{r}-\frac{f'}{2 f}\right)-H'_{tt}+H'_{yy}\right]-\frac{r^2 \omega H'_{tx}}{ f}=0\,,\\
   &k H'_{tx} \left(\frac{2}{r}-\frac{f'}{f}\right)+\omega (H_{xx}+H_{yy}) \left(\frac{1}{r}-\frac{f'}{2 f}\right)+k H'_{tx}+\omega \left(H'_{xx}+H'_{yy}\right)=0\,,\\
   &\frac{4}{r}H'_{tt}+\frac{2 \left(m-4 r^3+r \beta ^2\right)}{r \left(-2 m+2 r^3-r \beta ^2\right)}\left(H'_{xx}+H'_{yy}\right)+\frac{4 k^2}{2 m r-2 r^4+r^2 \beta ^2}(H_{tt}-H_{yy})\nonumber\\
   &-\frac{2 \left(2 m \beta ^2+r \beta ^4-2 r^3 \left(\beta ^2-2 \omega ^2\right)\right)}{r \left(2 m-2 r^3+r \beta ^2\right)^2}(H_{xx}+H_{yy})-\frac{16 k r^2 \omega }{\left(2 m-2 r^3+r \beta ^2\right)^2}H_{tx}=0\,.
 \end{align}
 The pure gauge solutions of these differential equations to the linear order in $\omega$ and $k$ yield
 \begin{eqnarray}
   H^{I}_{xt}=-\omega\,&,&\,H^{I}_{xx}=2k\,,\nonumber\\
   H^{II}_{tt}=2\omega\,&,&\,H^{II}_{xt}=kf\,,\nonumber\\
   H^{III}_{xx}=-8f^{1/2}\,&,&\,H^{III}_{yy}=-8f^{1/2}\,.
 \end{eqnarray}
 For the $\varepsilon=+1$ case, the incoming solutions of these differential equations to linear order in $\omega$ and $k$ are
  \begin{eqnarray}
   H^{inc}_{tt}&=&-\left(1-\frac{m}{r^3}-\frac{\beta ^2}{2r^2}\right)^{-\frac{i \omega }{6}}\left(\frac{24+\sqrt{2}}{12 \left(\sqrt{2}+\ln 3\right)} \beta ^2\right)\,,\nonumber\\
   H^{inc}_{xx}&=&-\left(1-\frac{m}{r^3}-\frac{\beta ^2}{2r^2}\right)^{-\frac{i \omega }{6}} \left(1-\frac{1}{6} i \omega  \ln\left[\frac{1}{3} \left(1+\frac{1}{r^2}+\frac{1}{r}\right)\right]-\frac{24-\ln 3}{12 \sqrt{2} \left(\sqrt{2}+\ln 3\right)} \beta ^2\right)\,,\nonumber\\
   H^{inc}_{yy}&=&\left(1-\frac{m}{r^3}-\frac{\beta ^2}{2r^2}\right)^{-\frac{i \omega }{6}} \left(1-\frac{1}{6} i \omega  \ln\left[\frac{1}{3} \left(1+\frac{1}{r^2}+\frac{1}{r}\right)\right]-\frac{24-\ln 3}{12 \sqrt{2} \left(\sqrt{2}+\ln 3\right)}\beta ^2\right)\,,\nonumber\\
   H^{inc}_{tx}&=&\left(1-\frac{m}{r^3}-\frac{\beta ^2}{2r^2}\right)^{-\frac{i \omega }{6}}\left(-\frac{1}{6} i k \left(1-\frac{1}{r^3}\right)\right)\,.
 \end{eqnarray}
 Then, the general solution of these differential equations is $H_{\mu\nu}=aH^{inc}_{\mu\nu}+bH^{I}_{\mu\nu}+cH^{II}_{\mu\nu}+dH^{III}_{\mu\nu}$ . The coefficients $a,b,c,d$ can be expressed in terms of the boundary field $H^0_{\mu\nu}=H_{\mu\nu}|_{r\rightarrow\infty}$, where $H^0_{\mu\nu}$ is $H^0_{tt},H^0_{s}=H^0_{xx}+H^0_{yy},H^0_{yy},H^0_{tx}$. We find $a,b,c,d$ are all proportional to
\begin{equation}
 \mathcal{F}^{-1}(\omega,k)=\bigg[-\frac{4}{3} k^2 \left(-4 i \omega +\frac{\left(24+\sqrt{2}\right)}{\sqrt{2}\beta ^2+\ln 3}\right)-\frac{2}{3} \left(k^2-2 \omega ^2\right) \left(24+4 i \omega  \ln 3+\frac{\sqrt{2}  (-24+\ln 3)}{\sqrt{2}+\ln 3}\beta ^2\right)\bigg]^{-1}\,.
\end{equation}
The location of the pole up to order of $\mathcal{O}(\omega^2,\omega k,k^2)$ is given by the equation $\mathcal{F}(\omega,k)=0$. The retarded Green's functions have the pole at
\begin{equation}
 \omega =\pm \frac{1}{\sqrt{2}}\left(1+\frac{\left(24+\sqrt{2}\right) \beta ^2}{24\left(\sqrt{2}+\ln 3\right)}\right)k- \frac{i}{12}\left(1+\beta ^2 \right)k^2\,,\label{dispersion1}
\end{equation}
corresponding to the speed of sound
\begin{equation} v_s(\varepsilon=1)=\frac{1}{\sqrt{2}}\left(1+\frac{\left(24+\sqrt{2}\right) \beta ^2}{24\left(\sqrt{2}+\ln 3\right)}\right)\,.\label{speed}
\end{equation}
When $\beta$ is strictly zero, the speed of sound is $1/\sqrt{2}$, exactly agreeing with the result of the conformal field theory.   When $\beta\neq 0$, Eq. (\ref{speed}) shows that sound-like modes exist in the range of short distance. According to what we have mentioned in the Introduction, the analysis suggests that breaking translational invariance makes the sound mode decay \cite{Davison:2014lua}, and we can look for the possible sound-like poles of the retarded Green function in an expansion when decay rate is small enough. The speed of sound (\ref{speed}) is shifted by $\beta$, and violates the bound $v_s^2\leq 1/p\ (p=2)$ for the 4D gravity dual theory. Indeed, there are several examples in holography where the bound can be violated \cite{Jokela:2015aha,Itsios:2016ffv,Hoyos:2016cob,Ecker:2017fyh}. For translational symmetry breaking systems, the longitudinal sound velocity can be modified by the charge-density-wave or spin-density-wave (see for example \cite{Varma}). The bound $v_s^2\leq 1/p\ (p=3)$ has been verified in several classes of strongly coupled theories with 5D gravity dual, such as the Sakai-Sugimoto model \cite{Benincasa:2006ei,Cai:2016sur}, the $D3=D7$ system \cite{Mateos:2007vn}, the single scalar model \cite{Cherman:2009tw,Hohler:2009tv} and the $N=2^*$ gauge theory \cite{Benincasa:2005iv}.  Therefore, the violation of the sound speed bound is worth future investigations. The sound attenuation $\frac{1+\beta^2}{12}$ is a monotonic function for $\epsilon=+1$. Actually, the imaginary part of $\omega$ is always negative in the $\epsilon=+1$ case, which indicates that quasinormal modes of the black hole are quite stable no matter how large $\beta$ are.

For the $\epsilon=-1$ case, the incoming solutions of these differential equations to the linear order in $\omega$ and $k$ are given by
  \begin{eqnarray}
   H^{inc}_{tt}&=&\left(1-\frac{m}{r^3}+\frac{\beta ^2}{2r^2}\right)^{-\frac{i \omega }{6}}\left(\frac{24+\sqrt{2}}{12 \left(\sqrt{2}+\ln 3\right)} \beta ^2\right)\,,\nonumber\\
   H^{inc}_{xx}&=&-\left(1-\frac{m}{r^3}+\frac{\beta ^2}{2r^2}\right)^{-\frac{i \omega }{6}} \left(1-\frac{1}{6} i \omega  \ln\left[\frac{1}{3} \left(1+\frac{1}{r^2}+\frac{1}{r}\right)\right]+\frac{24-\ln 3}{12 \sqrt{2} \left(\sqrt{2}+\ln 3\right)} \beta ^2\right)\,,\nonumber\\
   H^{inc}_{yy}&=&\left(1-\frac{m}{r^3}+\frac{\beta ^2}{2r^2}\right)^{-\frac{i \omega }{6}} \left(1-\frac{1}{6} i \omega  \ln\left[\frac{1}{3} \left(1+\frac{1}{r^2}+\frac{1}{r}\right)\right]+\frac{24-\ln 3}{12 \sqrt{2} \left(\sqrt{2}+\ln 3\right)}\beta ^2\right)\,,\nonumber\\
   H^{inc}_{tx}&=&\left(1-\frac{m}{r^3}+\frac{\beta ^2}{2r^2}\right)^{-\frac{i \omega }{6}}\left(-\frac{1}{6} i k \left(1-\frac{1}{r^3}\right)\right)\,.
 \end{eqnarray}
 Green's functions have the pole at
\begin{equation}
 \omega =\pm \frac{1}{\sqrt{2}}\left(1-\frac{\left(24+\sqrt{2}\right) \beta ^2}{24\left(\sqrt{2}+\ln 3\right)}\right)k- \frac{i}{12}\left(1-\beta ^2 \right)k^2\,.\label{dispersion2}
\end{equation}
 The corresponding sound velocity is
\begin{equation}
 v_s(\varepsilon=-1)=\frac{1}{\sqrt{2}}\left(1-\frac{\left(24+\sqrt{2}\right)\beta ^2}{24\left(\sqrt{2}+\ln 3\right)}\right).
 \end{equation} The sound attenuation in this case is $\frac{1-\beta^2}{12}$. The sound attenuation is negative when $\beta>1$. Whenever $\varepsilon=\pm1$, the speed of the sound wave could be definitely able to recover $1/\sqrt{2}$ in the small $\beta$ limit \cite{Herzog:2003ke}.

To investigate stability in the previous paragraph, one can further study the null energy condition. The averaged null energy condition states that for any integral curve
 of the null vector field $K^a$,  we must have $\int T_{ab}K^a K^b\geq 0 $ or equivalently $T^{tt}+T^{x_i x_i}\geq 0$, where $T_{ab}$ is the stress tensor. Since the Hawking temperature must be non-negative $T \geq 0$, this gives a constraint for $\beta$. And together with the location of the event horizon, $\beta$ is less than $\sqrt{6}m^{1/3}$. Then the null energy condition is satisfied $T^{tt}+T^{x_i x_i}>0$  for any $r$ range from $r_H$ to infinity. Therefore, the null energy condition is satisfied  for the $\varepsilon=1$ case. However, if $\varepsilon=-1$ and $\beta\neq0$, the imaginary part of $\omega$ may be positive and signals the dynamical instability of the black hole. The null energy condition is violated $T^{tt}+T^{x_i x_i}\rightarrow-2\beta^2<0$ as $r\rightarrow\infty$. The possible reason for such a dynamical instability situation may be that $\varepsilon=-1$ corresponds to the phantom case.
 In order to examine the impacts of axion on the stability of the black hole, numerical computations of the quasinormal modes corresponding to equations (\ref{f1}-\ref{f5}) are presented in Fig. 1. We first consider the $\beta=0$ case, which reduces to the stable Schwarzschild-AdS case.  When $\beta\neq 0$ and $\varepsilon =1$, the black hole receives corrections from linear axions but it is still stable as shown in (b) of Fig. 1. As to the phantom-like case $\beta\neq 0$  and $\varepsilon =-1$, shown in (c) and (d) of Fig. 1,  the imaginary part of the quasinormal modes becomes positive as $\beta>1$, indicating instabilities of the black hole. However, as $\beta< 1$, the unstable modes do not appear. These numerical results are in fact consistent with (\ref{dispersion1}) and (\ref{dispersion2}). Note  that we have taken the coordinate transformation $z=r_H/r$, and the horizon condition gives a constraint for $m=r^3_H-r^2_H\beta^2/2$. Without losing generality, we set $r_H=1$, inserted the dimensionless frequency and momentum $\bm{\omega}=\omega/(2\pi T)\,,\,\bm{k}=k/(2\pi T)$ and imposed the incoming condition with $f(r)^{-\frac{i}{2}\omega}$.

\begin{figure}[!t]
\begin{minipage}{0.48\linewidth}
\centerline{\includegraphics[width=7.0cm]{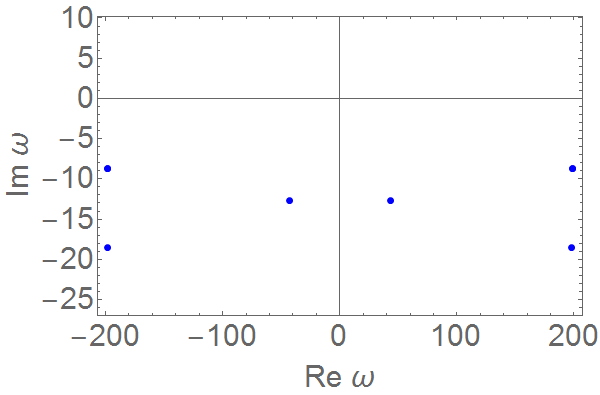}}
\centerline{(a)}
\end{minipage}
\hfill
\begin{minipage}{0.48\linewidth}
\centerline{\includegraphics[width=7.0cm]{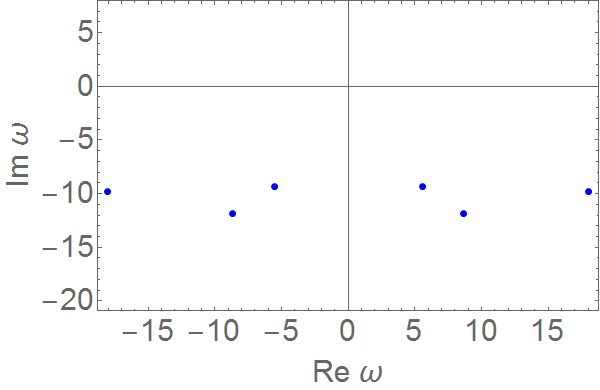}}
\centerline{(b)}
\end{minipage}
\vfill
\begin{minipage}{0.48\linewidth}
\centerline{\includegraphics[width=7.0cm]{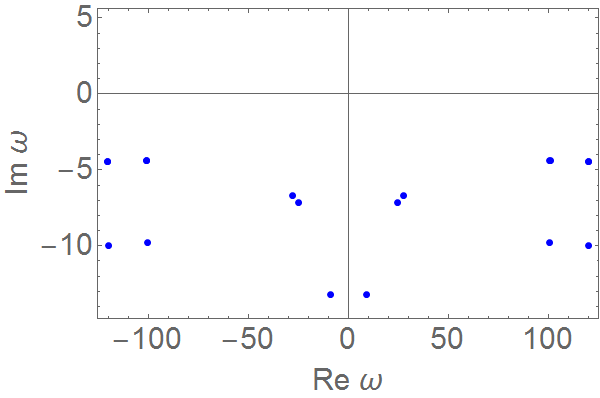}}
\centerline{(c)}
\end{minipage}
\hfill
\begin{minipage}{0.48\linewidth}
\centerline{\includegraphics[width=7.0cm]{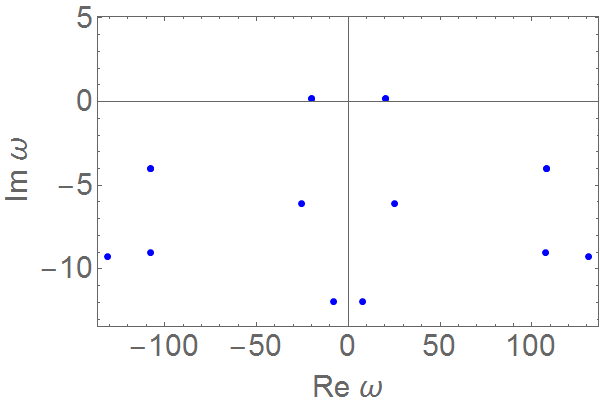}}
\centerline{(d)}
\end{minipage}
\caption{\label{fig:Figure 1} The quasinormal modes for the fluctuations of the
metric and axion field with different overtone numbers n in the complex plane. (a) The quasinormal modes with $\epsilon=\pm 1$ and $\beta=0$. (b) The quasinormal modes with $\epsilon=+1$ and $\beta=1.05$. (c): The imaginary part of quasinormal modes are negative with $\epsilon=-1$ and $\beta=0.95$. (d): A pair of poles occurs in the upper half plane when $\epsilon=-1$ and $\beta=1.05$. Here we have neglected overdamped modes.}
\end{figure}

\section{Hyperscaling violating black hole solutions}
In \cite{Ge:2016sel}, a black hole solution in Einstein-Maxwell-axion-dilaton theory with a hyperscaling violation exponent was obtained. This black hole solution is not really an asymptotic AdS solution and in principle can be interpreted as an IR geometry embedded in the AdS space. The linear-temperature-dependence resistivity and quadratic-temperature-dependence inverse Hall angle can be achieved through the gauge-gravity duality (see also \cite{li2017,Blauvelt:2017koq,Cremonini:2018kla,lili2017} for related references). Here, we study the stability of the black hole in the case that the null energy condition is violated. The Einstein-Maxwell-dilaton theory with additional axion fields, which is used for breaking translational invariance in the boundary theory, is given by \cite{Ge:2016sel}
\be
S=\int d^{4}x \sqrt{-g} \bigg[\bigg(R+V(\phi)+\frac{1}{2}\partial \phi^2-\frac{1}{2}Y(\phi)\sum^{2}_{i=1}\partial\chi^2_{i}\bigg)-\frac{1}{4}Z(\phi)F^2\bigg].\label{BH}
\ee
where we selected $16 \pi G=g^2=L=k_B=e=1$, where $L$ is the AdS radius, $g^2$ is the $3+1$-dimensional gauge coupling constant, and $G$ is the Newton's constant.

The full black hole solution is given by
\bea\label{metric}
ds^2&=&-g_{tt}dt^2+g_{rr}dr^2+g_{xx}(dx^2+dy^2)\nonumber\\&=&\!r^{-\theta}\bigg[-r^2f(r)dt^2+\frac{dr^2}{r^2f(r)}+r^2(dx^2+dy^2)\bigg]\!,\nonumber\\
f(r)&=&1-\frac{m}{r^2}-\frac{q^2 \ln r}{2 r^2}-\frac{\beta^2}{r}+\frac{B^2}{8r^4},\\
 A&=&q \ln r dt+\frac{B}{2}(xdy-ydx), e^{\phi}=r,  \chi_i=\beta\delta_{ia} x^a, \nonumber\\ V(\phi)&=&2r+\mathcal{C}r^{-3}, Z(\phi)=r^{-1}, Y(\phi)=r, \nonumber
\eea
where $\theta$, $m$, $q$, and $B$ are the parameters related to the hyperscaling violation factor, mass, charge, and magnetic field, respectively.
Solving the equations of motion, we find that only when the constant $\mathcal{C}$ takes the form $\mathcal{C}(r)=B^2 /8$, the Einstein equation is satisfied.
It is intriguing to see that the magnetic field $B$ appears in the potential $V(\phi)$, and this is because we only consider the IR geometry so that the magnetic field is fixed in the action. The boundary is located at $r\rightarrow +\infty$ and the nondegenerate horizon is located at $r=r_H$ where $f(r_H)=0$. The Hawking temperature is given by $T=\frac{r_H}{2\pi}\big(1-\frac{q^2}{4r^2_H}-\frac{\beta^2}{2r_H}-\frac{B^2}{8r^4_H}\big)$ and the entropy density is $S=4\pi r_H$.

\section{Tensor-type perturbations}
 Now the problem is that a positive kinetic term of the dilaton in the action looks like a phantom term. In our case the dilaton field $\phi$ only depends on the radial coordinate.
 Next, let us investigate whether a positive kinetic term of dilaton leads to pathological boundary field theory by exploring the causality analysis, since one may worry that the violation of the null energy condition could result in nonunitary deformation and negative field couplings. Considered the tensor type of perturbation of the form $h_{xy}=e^{-i\omega t+i k_x x}h_{xy}(r)$, the equation of motion for $h_{xy}$ is given by
 \be
 \partial_{r}\bigg(\mathcal{N}^{ry}\partial_{r} h^x_{y}\bigg)-k^2_x \mathcal{N}^{xy}h^x_{y}-\omega^2 \mathcal{N}^{ty}h^x_{y}=0,
 \ee
 with the notation $\mathcal{N}^{\mu\nu}=\frac{1}{2}g_{xx}\sqrt{-g}g^{\mu\mu}g^{\nu\nu}$. In order to check the causality on the boundary, we simply assume
 \be
 h^x_{y}=e^{-i\omega t+i k_x x+ik_r r}.
 \ee
In the large-momentum limit, the effective geodesic equation can be rewritten as $k^{\mu}k^{\nu}g^{\rm{eff}}_{\mu\nu}=0$, from which we can read off the effective metric as follows
\be
d{s}^2_{\rm{eff}}=f(r)\bigg(-dt^2+\frac{1}{f(r)}dx^2\bigg)+\frac{1}{r^4 f(r)}dr^2.
\ee
The local speed of light is then given by
\be
c^2_g=f(r).
\ee
In the standard Ferrerman-Graham coordinate $v=1/r\rightarrow 0$, the expansion of the function can be written as
\be
f=1-\beta^2 v-v^2 \left(\frac{1}{2} q^2 \log (v)-m\right)+\mathcal{O}(v^4).
\ee
We can expand the local speed of light near the boundary $v \rightarrow 0$
\be
c^2_g-1=-\beta^2 v+\mathcal{O}(v^2).
\ee
Note that the local speed of gravitons should be smaller than 1 (the speed of the boundary conformal field theory) and this requires $c^2_g -1 <0$.
This then leads to the condition $\beta^2>0$. This requirement is automatically satisfied by the value of $\beta$ obtained in the black hole solution \footnote{Note that we did not study the tensor type perturbations in the previous section since some related discussions have already been done in \cite{wang}.}.

As it was proved in \cite{causality,gsl,gb1,gb2}, the group velocity of the graviton is given by
\be
v_g=\frac{\Delta x}{\Delta t}\sim c_g.
\ee
That is to say, as near the boundary $c_g$  is smaller than 1, the
propagation of signals in the boundary theory with speed $\frac{\Delta x}{\Delta t}$ cannot be superluminal.

\section{Sound modes}
For simplicity of the calculations of sound modes, we consider a neutral case of black hole solutions given in (\ref{metric}) with $q=0$ and $B=0$ in what follows.   The  black hole solution becomes
\bea\label{metric}
ds^2&=&-g_{tt}dt^2+g_{rr}dr^2+g_{xx}(dx^2+dy^2)\nonumber\\&=&\!r^{-\theta}\bigg[-r^2f(r)dt^2+\frac{dr^2}{r^2f(r)}+r^2(dx^2+dy^2)\bigg]\!,\nonumber\\
f(r)&=&1-\frac{m}{r^2}-\frac{\beta^2}{r},\\
e^{\phi}&=&r,  \chi_i=\beta \delta_{ia} x^a, \nonumber\\ V(\phi)&=&2r, Z(\phi)=r^{-1}, Y(\phi)=r, \nonumber
\eea
Inserting the similar definitions of the gauge and fluctuations as in Sec. II,
\bea
h_{tt}&=&-e^{-i\omega t+i kx}g_{tt}H_{tt},~~~ h_{tx}=h_{xt}=e^{-i\omega t+i kx}g_{xx}H_{tx},\\
h_{xx}&=&e^{-i\omega t+i kx}g_{xx}H_{xx}, ~~~ h_{yy}=e^{-i\omega t+i kx}g_{yy}H_{yy},\\
\delta \phi&=&e^{-i\omega t+i kx}\mathcal{P}(r), ~~~~\delta \chi_i=e^{-i\omega t+i kx}\mathcal{X}_i (r),
\eea
to the equations of motion for the action (\ref{BH}), we expand all the equations of motion at a linearized level as
\bea
&&H''_{tt}+\ln'\bigg(\frac{g_{tt}g_{xx}}{g^{1/2}_{rr}}\bigg)H'_{tt}-\ln'(g^{1/2}_{tt})H'_{ii}-g_{rr}\bigg(\frac{\omega^2}{g_{tt}}H_{ii}+\frac{k^2}{g_{xx}}H_{tt}+2\frac{\omega k}{g_{tt}}H_{tx}\bigg)\nonumber\\
&&+2rg_{rr}\mathcal{P}=0,\label{EOM1}\\
&&H''_{tx}+\ln'\bigg(\frac{g^2_{xx}}{g^{1/2}_{rr}g^{1/2}_{tt}}\bigg)H'_{tx}+\frac{g_{rr}}{g_{xx}}\omega k H_{aa}+h(\beta)H_{tx}-i\beta \omega r \frac{g_{rr}}{g_{xx}}\mathcal{X}_1=0,\label{EOM2}\\
&& H''_{xx}+\ln'\bigg(\frac{g^{1/2}_{tt}g^{3/2}_{xx}}{g^{1/2}_{rr}}\bigg)H'_{xx}+\ln'(g^{1/2}_{xx})\bigg(H'_{yy}-H'_{tt}\bigg)\nonumber\\
\eea
\bea
&&+g_{rr}\bigg(\frac{\omega^2}{g_{tt}}H_{xx}+\frac{k^2}{g_{xx}}(H_{tt}-H_{yy})+2\frac{\omega k}{g_{tt}}H_{tx}\bigg)+h(\beta)H_{xx}-(2-\frac{\beta^2}{g_{xx}})rg_{rr}\mathcal{P}\nonumber\\
&&+2ik\beta r \frac{g_{rr}}{g_{xx}}\mathcal{X}_1=0,\label{EOM3}\\
&& H''_{yy}+\ln'\bigg(\frac{g^{1/2}_{tt}g^{3/2}_{xx}}{g^{1/2}_{rr}}\bigg)H'_{yy}+\ln'(g^{1/2}_{xx})\bigg(H'_{xx}-H'_{tt}\bigg)+g_{rr}\bigg(\frac{\omega^2}{g_{tt}}-\frac{k^2}{g_{xx}}\bigg)H_{yy}\nonumber\\
&&+h(\beta)H_{yy}-
\bigg(2-\frac{\beta^2}{g_{xx}}\bigg)g_{rr}r\mathcal{P}=0,\label{EOM4}\\
&& \mathcal{P}''+\ln'\bigg(\frac{g^{1/2}_{tt}g_{xx}}{g^{1/2}_{rr}}\bigg) \mathcal{P}'+g_{rr}\bigg(\frac{\omega^2}{g_{tt}}-\frac{k^2}{g_{xx}}+\beta^2 r \frac{1}{g_{xx}}\bigg)\mathcal{P}+\frac{1}{2}\phi'(H_{ii}-H_{tt})'
\nonumber\\
&&-\frac{\beta^2}{2}r\frac{g_{rr}}{g_{xx}}H_{ii}-i\beta k r\frac{g_{rr}}{g_{xx}}\mathcal{X}_1-2rg_{rr}\mathcal{P}=0,\label{EOM5}\\
&&\mathcal{X}''_1+\bigg(\frac{1}{r}+\ln\frac{g^{1/2}_{tt}g_{xx}}{g^{1/2}_{rr}}\bigg)\mathcal{X}'_1+g_{rr}\bigg(\frac{\omega^2}{g_{tt}}-\frac{k^2}{g_{xx}}\bigg)\mathcal{X}_1
\nonumber\\
&&-i\beta g_{rr}\bigg[\frac{k}{2g_{xx}}H_{tt}+\frac{\omega}{g_{tt}}H_{tx}+\frac{k}{2 g_{xx}}(H_{xx}-H_{yy})\bigg]=0,\label{EOM6}
\eea
where $h(\beta)=-\beta^2 g_{rr}$. Note that we take $q=0$ and $B=0$ in (\ref{EOM1})-(\ref{EOM6}). The presence of the $U(1)$ field makes the calculation of the sound speed more complicated and more tedious and usually the fluctuations of gauge fields do not modify the sound velocity (see \cite{matsuo} for an example).

The constraint equations associated with the gauge fixing condition $h_{\mu r}=0$ are
\bea
&&H'_{ii}+\ln'\bigg(\frac{g^{1/2}_{xx}}{g^{1/2}_{tt}}\bigg)H_{ii}+\frac{k}{\omega}H'_{tx}+2\frac{k}{\omega}\ln'\bigg(\frac{g^{1/2}_{xx}}{g^{1/2}_{tt}}\bigg)H_{tx}-\phi'\mathcal{P}=0,\\
&&H'_{tt}-\ln'\bigg(\frac{g^{1/2}_{xx}}{g^{1/2}_{tt}}\bigg)H_{tt}+\frac{\omega}{k}\frac{g_{xx}}{g_{tt}}H'_{tx}-H'_{yy}+\phi'\mathcal{P}+i\beta\frac{r}{q}\chi'_1=0,\\
&&\ln'\bigg(g^{1/2}_{xx}g^{1/2}_{tt}\bigg)H'_{ii}-\ln'(g_{xx})H'_{tt}+g_{rr}\bigg[\frac{\omega^2}{g_{tt}}H_{xx}+\frac{k^2}{g_{xx}}(H_{tt}-H_{yy})+2\frac{\omega k}{g_{tt}}H_{tx}\bigg]
\nonumber\\
&&+\phi'\mathcal{P}'-2rg_{rr}\mathcal{P}+i\beta k r\frac{g_{rr}}{g_{xx}}\mathcal{X}_1+\beta^2 r g_{rr}\mathcal{P}+\frac{1}{2}h(\beta)H_{ii}=0.
\eea
One may note that there is still a residual gauge freedom under the infinitesimal diffeomorphism
\bea
&& x^{\mu}\rightarrow x^{\mu}+\xi^{\mu},\nonumber\\
&& \delta g_{\mu\nu}\rightarrow \delta g_{\mu\nu}-\nabla_{\mu}\xi_{\nu}-\nabla_{\nu}\xi_{\mu},\\
&& \delta \phi\rightarrow \delta \phi-\partial^{\mu}\phi \xi_{\mu},\nonumber\\
&& \delta \chi_i \rightarrow \delta \chi_i-\partial^{\mu}\chi \xi_{\mu},
\eea
where $\xi_{\mu}=\xi_{\mu}(r)e^{-i\omega t+i kx}$. We can check that the following combinations are a set of gauge invariant fluctuations
\bea
Z_H &=&2 k \omega H_{tx}+\omega^2 H_{xx}+k^2 \frac{g_{tt}}{g_{xx}}H_{tt}+\bigg(k^2\frac{\ln'(g^{1/2}_{tt})g_{tt}}{\ln'(g^{1/2}_{xx})g_{xx}}-\omega^2\bigg)H_{yy},\\
Z_{\phi} &=&\mathcal{P}-\frac{\phi'}{\ln'(g_{xx})}H_{yy},\\
Z_{\chi} &=&\mathcal{X}_1-\frac{i}{2k}\frac{\beta\phi'}{\ln'(g_{xx})}(H_{xx}-H_{yy}).
\eea
It is convenient to deal with fluctuations by introducing $Z_H,Z_{\phi},Z_{\chi}$ \cite{Buchel:2005cv}. $Z_H$ is the gauge-invariant combination linear in gravitational fluctuations of the sound channel : $h_{tt},h_{tx},h_{xx},h_{yy}$.
Unfortunately, we are not able to totally decouple the equations of motion in the presence of the linear axion fields. The resulting master equations  are given in the Appendix
\bea
&&Z''_{\phi}+\frac{m+(2\beta^2-3r)r}{r[m+(\beta^2-r)r]}Z'_{\phi}+\frac{mk^2+r[(\beta^2-r)k^2+r\omega^2 r]}{r^2[m+(\beta^2-r)r]^2}Z_{\phi}\nonumber\\
&&+\frac{ik\beta}{r[m+(\beta^2-r)r]}Z_{\chi}=0\,,\label{MEQ1}\\
&&Z''_{H}+\mathcal{F}(r)Z'_H+\mathcal{G}(r)Z_H+\mathcal{H}(r)Z_{\phi}=\mathcal{K}(r)\chi'_1+\mathcal{S}(r)H_{yy}\,,\label{MEQ2}\\
&&Z''_{\chi}+\frac{m+(2\beta^2-2r)r}{r[m+(\beta^2-r)r]}Z'_{\chi}+\frac{m(k^2-\beta^2 r)+r[k^2(\beta^2-r)+r(\beta^2 r-\beta^4+\omega^2)]}{r^2[m+(\beta^2-r)r]^2}Z_{\chi}
\nonumber\\
&&-\frac{i\beta}{k[m+(\beta^2-r)r]^2}Z_H=0\,.\label{MEQ3}
\eea
  In the $\beta=0$ case, the axion fields and the equation for $Z_{\chi}$ do not appear. Under this condition, the master equations reduce to two decoupled equations for $Z_{\phi}$ and $Z_{H}$ as given in \cite{son,spring,tarrio}. We can impose the ingoing boundary as follows
\bea
&&Z_{\phi}=f(r)^{-\frac{i\omega}{4\pi T}}\bigg(\mathcal{Y}_0(r)+k \mathcal{Y}_1 (r)+\ldots \bigg),\nonumber\\
&&Z_H=f(r)^{-\frac{i\omega}{4\pi T}}\bigg({Y}_0(r)+k {Y}_1 (r)+\ldots \bigg),\nonumber\\
&& \omega(k)=\omega_1 k+\omega_2 k^2+\ldots.
\eea
Inserting this ansatz into (\rm{A1}), expanding the result in powers of $k$ and neglecting terms of $\mathcal{O}(k^2)$ and higher, we find that the only nonsingular solution for $Z_{\phi}$ is a constant, which we set to zero by the boundary conditions at infinity. Inserting now $Z_{\phi}=0$ into the main equation for $Z_H$ and solving perturbatively with  the Dirichlet boundary condition as given in \cite{son,spring,tarrio}
\bea
Z_H(r)\bigg|_{r=\infty}=0,\label{bc1}
\eea
we obtain the expression for $\omega(k)$ as follows
\bea
\omega=k+\mathcal{O}(k^2).
\eea
We finally obtain the sound velocity as
\be\label{velocity}
v_s=1.
\ee
This indicates that the sound velocity obtained here coincides with the sound velocity of $2+1$-dimensional BTZ black holes. There exists a class of charged BTZ-like black hole solutions in Lifshitz spacetime with a hyperscaling violating factor \cite{Ge:2017fix}.

As a double check, we follow \cite{spring}, notice that the condition $R^t_t=R^x_x$ is satisfied for our case and expand $-g_{tt}$ as
\be
-g_{tt}=\frac{a_0}{a_2-p}g_{xx}+a_1 g^{a_2-p+1}_{xx},
\ee
where $a_0$, $a_1$ and $a_2$ are constants independent of $r$ and $p$ is the number of the spatial dimensions. Now we return to the equation for $Z_H$ and go through the same steps of inserting the incoming wave condition. Expanding in powers of $k$ and applying the boundary conditions
leads to the dispersion relation
\be
\omega(k)=\sqrt{\frac{a_0-a_2}{p}}k+\mathcal{O}(k^2).
\ee
Comparing this dispersion relation with the expected hydrodynamics dispersion relation, we obtain the speed of sound \cite{spring}
\be
v_s=\sqrt{\frac{a_0-a_2}{p}}.
\ee
In our case, $a_0=2$, $a_2=0$ and $p=2$. Therefore the speed of sound reads $v_s=1$.

So far, the results are obtained in the $\beta\rightarrow 0$ limit. Since we are not able to decouple the equations of motion in the absence of the momentum dissipation, we would like to consider solving this problem numerically. Moreover, we have not considered the shear viscosity and bulk viscosity in this setup. It was found in \cite{hartnoll,wang} that the shear viscosity to entropy density ratio can be greatly modified in the presence of the momentum dissipation in the AdS space. We expect that the bulk viscosity could also receive great modifications in the presence of the momentum dissipation term.

Although the explicit form of the sound velocity $v_s$ could not be evaluated, the contribution of $\beta$ to $v_s$ still could be discussed briefly. ${Y}_0(r)$ and ${Y}_1 (r)$ are regular at the horizon. Accordingly, without the loss of generality, the boundary condition at the horizon
\be
f(r)^{\frac{-i\omega}{4\pi T}}Z_H(r)\bigg|_{r=\frac{1}{2}\big(\beta^2+\sqrt{4m+\beta^4}\big)}=1.\label{bc2}
\ee
Imposing the two boundary conditions (\ref{bc1}), (\ref{bc2}) and expanding with small $k$ and $\beta$, we obtain the sound velocity is 1 for the leading order in $\mathcal{O}(k^0)$. When $\beta=0$, $v_s$ could return to (\ref{velocity}). For the next-to-leading order, this approach is not enough to determine the contribution of $\beta$ to the sound velocity.

In order to see whether the phantom-like term $\partial\phi^2$ introduces any instabilities, we compute the quasinormal modes numerically for the sound channel in leading order $\mathcal{O}(k)$ \cite{Bronnikov:2012,kuang2017,Kodama:2003kk,Ge:2008ni,Kaminski:2009ce}. We could further rewrite the expressions in terms of the dimensionless frequency and momentum $\bm{\omega}=\omega/(2\pi T)\,,\,\bm{k}=k/(2\pi T)$ where $T=\frac{r_H}{2\pi}(1-\frac{\beta^2}{2r_H})$ is Hawking temperature and $r_H$ is the horizon radius. We take $u=r_H/r$ and further require the incoming condition $Z_H(u)=\mathcal{Z}_H(u)f(u)^{-\frac{1}{2} (i \bm{\omega} )}\,,\,Z_\phi(u)=\mathcal{Z}_\phi(u)f(u)^{-\frac{1}{2} (i \bm{\omega} )}$ and $Z_\chi(u)=\mathcal{Z}_\chi(u)f(u)^{-\frac{1}{2} (i \bm{\omega} )}$.

The quasinormal frequencies with $\bm{\omega}\,,\,\bm{k}$ are plotted in Fig. 2. Note that $\bm{\omega}=\omega\,,\,\bm{k}=k$ when $\beta=0$. Remarkably, as $\beta=0$, the numerical calculation yields the real part of the quasinormal frequency ${\rm Re}~\bm{\omega}=\pm \bm{k}$, inferring the sound velocity is $v_s=1$ which is agreeing with (59). In fact, we have checked that the quasinormal modes change little when $\bm{\omega}\rightarrow \omega\,,\,\bm{k}\rightarrow k$. In this sense, we could still use $\omega$ and $k$ in the previous sections. As $\beta\neq 0$, the imaginary part of quasinormal frequency becomes positive for different $n$ as shown in Fig. 2. This result signalizes that the background black hole is unstable.
\begin{figure}[!t]
\begin{centering}
\includegraphics[scale=0.37]{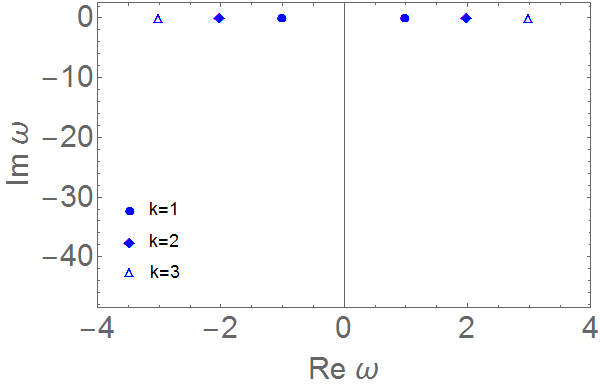}
\includegraphics[scale=0.37]{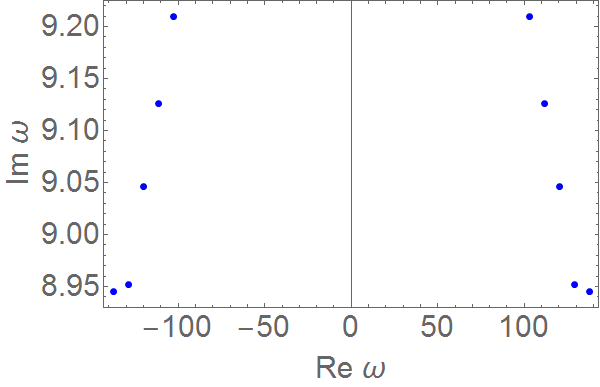}
\par\end{centering}
\caption{\label{fig:Figure 2} Plots of the sound-like poles by introducing gauge invariant master field $Z_H\,,\,Z_\phi\,,\,Z_\chi$. Left: The sound-like quasinormal spectrum with various $\bm{k}$ at fixed $\beta=0$. Three pairs of poles demonstrate the dispersion relation $\bm{\omega}=\pm \bm{k}$.  Notice that the $\mathcal{K}(r)\chi'_1+\mathcal{S}(r)H_{yy}$ term in (\ref{MEQ2}) vanishes if $\beta=0$. As a result, the master equations (\ref{MEQ1})-(\ref{MEQ3}) can be solved together. Right: The sound-like quasinormal spectrum at $\beta=0.1$ and $\bm{k}=1$. Up to the leading order in $\bm{k}$ expansion, the fluctuation equation (\ref{MEQ2}) is decoupled. The points from top to bottom correspond to $n=0$ to $n=6$, respectively.}
\end{figure}

\section{Discussion and conclusion}
In this paper, the sound modes of momentum dissipated holographic systems and their stability are investigated. We focus on two holographic models, and investigate the sound velocity, the quasinormal modes of scalar-type metric perturbation and the dispersion relation. Concerning the observational dark energy in the Universe, we study phantom-like terms and their influences on the dynamical stability of holographic black holes.

We first study the Einstein-linear axion model, and calculate the sound velocity and attenuation. We first develop a new analytical method for this model, which is different from \cite{Davison:2014lua}, and solve the coupled equations of the scalar-type perturbation without exploring the master field equation. But the result is comparable to that of \cite{Davison:2014lua} and in a certain limit can recover the result given in \cite{Davison:2014lua}. Moreover, we also present our numerical results in Fig. 1.  Our numerical result suggests that phantom-like terms might lead to dynamical instability. At the qualitative level, the analytical and numerical results agree with each other very well. This numerical method is effective to compute fluctuations equations which are actually not decoupled.

We then study Einstein-Maxwell-dilaton-axion model, including the dynamical stability of the hyperscaling violating black hole solutions. The tensor-type perturbation and causality are discussed in this black hole background. No causality violation happens. The sound velocity and quasinormal modes of the sound modes have been studied in this background. We analytically obtain the dispersion relation from master equations up to the leading order in $k$ expansion and the leading order in $\beta$ expansion. The analytical result matches with the exact numerical result from master equations at fixed $\beta=0$. In the same way, we obtain the quasinormal frequency with $\beta\neq 0$. According to the sound-like poles with $\beta\neq 0$ in Fig. 2, we conclude that an instability driven by phantom-like term exists in the hyperscaling violating black hole. The instability and the unstable quasinormal modes found here may be related to the chaos behavior found in \cite{Grozdanov:2017ajz}. Frequency and momentum follow from a dispersion relation of a hydrodynamic sound mode give holographic Lyapunov exponent $\lambda_L$ and the butterfly velocity $v_B$ as in \cite{Grozdanov:2017ajz}. The momentum is conserved in this holographic system. So it is natural to ask what about the momentum dissipated case? We leave it for future study.

\begin{center}
{\bf ACKNOWLEDGEMENTS}
\end{center}
We would like to thank Richard A. Davison, Akihiro Ishibashi, Li Li and Wei-Jia Li and  for valuable comments. The authors were partly supported by NSFC, China (No.11875184 $\&$ No.11805117).

\section*{Appendix: THE MASTER EQUATIONS}
In this appendix, we collect the master equations for the fluctuations of momentum relaxed theories in Sec. V.
\bea
&&Z''_{\phi}+\frac{m+(2\beta^2-3r)r}{r[m+(\beta^2-r)r]}Z'_{\phi}+\frac{mk^2+r[(\beta^2-r)k^2+r\omega^2 r]}{r^2[m+(\beta^2-r)r]^2}Z_{\phi}\nonumber\\
&&+\frac{ik\beta}{r[m+(\beta^2-r)r]}Z_{\chi}=0\,,\nonumber
\eea
\be
Z''_{H}+\mathcal{F}(r)Z'_H+\mathcal{G}(r)Z_H+\mathcal{H}(r)Z_{\phi}=\mathcal{K}(r)\chi'_1+\mathcal{S}(r)H_{yy}\,,\nonumber\tag{A1}
\ee
\bea
&&Z''_{\chi}+\frac{m+(2\beta^2-2r)r}{r[m+(\beta^2-r)r]}Z'_{\chi}+\frac{m(k^2-\beta^2 r)+r[k^2(\beta^2-r)+r(\beta^2 r-\beta^4+\omega^2)]}{r^2[m+(\beta^2-r)r]^2}Z_{\chi}
\nonumber\\
&&-\frac{i\beta}{k[m+(\beta^2-r)r]^2}Z_H=0\,,\nonumber
\eea
where
\bea
&&\mathcal{F}(r)=\frac{m \left(q^2 \left(3 \beta ^2-2 r\right)+2 r \omega ^2\right)+r \left(q^2 \left(4 \beta ^4+6 r^2-9 \beta ^2 r\right)+2 r \omega ^2 \left(2 \beta ^2-3 r\right)\right)}{r \left(r \left(r-\beta ^2\right)-m\right) \left(q^2 \left(2 r-\beta ^2\right)-2 r \omega ^2\right)},\nonumber\\
&&\mathcal{G}(r)=\bigg[m \left(q^4 \left(2 r-\beta ^2\right)+2 q^2 r \left(-2 \beta ^4+2 \beta ^2 r-\omega ^2\right)-2 \beta ^2 r^2 \omega ^2\right)-2 \beta ^2 m^2 q^2-r \nonumber\\
&&\left(q^4 \left(\beta ^4+2 r^2-3 \beta ^2 r\right)+q^2 r \left(2 \beta ^6+3 \beta ^2 \omega ^2+2 \beta ^2 r^2-4 r \left(\beta ^4+\omega ^2\right)\right)+2 r^2 \omega ^2 \left(\beta ^4-\beta ^2 r+\omega ^2\right)\right)\bigg]\nonumber\\
&&\bigg[r^2 \left(m+r \left(\beta ^2-r\right)\right)^2 \left(q^2 \left(2 r-\beta ^2\right)-2 r \omega ^2\right)\bigg]^{-1},\nonumber\\
&&\mathcal{H}(r)=-\frac{q^2 \left(2 m+\beta ^2 r\right) \left(m \left(q^2 \left(4 r-\beta ^2\right)-4 r \omega ^2\right)+\beta ^2 r^2 \left(q^2-2 \omega ^2\right)\right)}{r^4 \left(r \left(r-\beta ^2\right)-m\right) \left(q^2 \left(2 r-\beta ^2\right)-2 r \omega ^2\right)},\nonumber\\
&&\mathcal{K}(r)=-\frac{2 i \beta  q \left(m+r \left(\beta ^2-r\right)\right) \left(m \left(q^2 \left(\beta ^2-4 r\right)+4 r \omega ^2\right)-\beta ^2 r^2 \left(q^2-2 \omega ^2\right)\right)}{r^2 \left(r \left(r-\beta ^2\right)-m\right) \left(q^2 \left(2 r-\beta ^2\right)-2 r \omega ^2\right)},\nonumber\\
&&\mathcal{S}(r)=-\frac{2 \beta ^2 q^2 \left(m \left(q^2 \left(\beta ^2-4 r\right)+4 r \omega ^2\right)-\beta ^2 r^2 \left(q^2-2 \omega ^2\right)\right)}{r^3 \left(r \left(r-\beta ^2\right)-m\right) \left(q^2 \left(2 r-\beta ^2\right)-2 r \omega ^2\right)}.\nonumber
\eea

\end{document}